\documentclass[aps,prd,10pt,twocolumn,superscriptaddress,floatfix]{revtex4-1}

\usepackage{amsmath} 
\usepackage{graphicx}
\usepackage{color}

\newcommand{\Tr}{\mathrm{Tr}}
\newcommand{\intd}{\textrm{d}}

\begin{document}
           

\title{Supernova Equation of State with an extended SU(3) Quark-Meson Model}

\author{Thomas~Beisitzer}\email{beisitzer@thphys.uni-heidelberg.de}
\affiliation{Institut f\"{u}r Theoretische Physik, Universit\"at Heidelberg,
Philosophenweg 16, D-69120, Heidelberg, Germany}

\author{Rainer~Stiele}\email{r.stiele@thphys.uni-heidelberg.de}
\affiliation{Institut f\"{u}r Theoretische Physik, Universit\"at Heidelberg,
Philosophenweg 16, D-69120, Heidelberg, Germany}

\author{J\"urgen~Schaffner-Bielich}\email{schaffner@astro.uni-frankfurt.de}
\affiliation{Institut f\"{u}r Theoretische Physik, Universit\"at Heidelberg,
Philosophenweg 16, D-69120, Heidelberg, Germany}
\affiliation{{Institut f\"{u}r Theoretische Physik, Goethe-Universit\"at Frankfurt,
Max-von-Laue-Stra\ss e 1, D-60438, Frankfurt am Main, Germany}}
\affiliation{ExtreMe Matter Institute EMMI, GSI Helmholtzzentrum f\"ur
Schwerionenforschung GmbH, Planckstra\ss e 1, D-64291 Darmstadt, Germany}


\begin{abstract} 
  The quark-meson model is investigated for the two- and three-flavor case
  extended by contributions of vector mesons under conditions encountered in
  core-collapse supernova matter. Typical temperature ranges, densities and
  electron fractions, as found in core-collapse supernova simulations, are
  studied by implementing charge neutrality and local $\beta$-equilibrium with
  respect to weak interactions. Within this framework, we analyze the
  resulting phase diagram and equation of state (EoS) and investigate the
  impact of undetermined parameters of the model. The EoS turns out to be
  relatively independent on the entropy per baryon but there are significant
  changes when going from the two-flavor to the three-flavor case due to the
  nontrivial contribution from the strange quarks which stay massive even at
  high densities. While an increasing vector meson coupling constant leads to
  a substantial stiffening of the EoS, we find that the impact of changing the
  scalar meson mass is equally strong and results in a softening of the EoS
  for increasing values.
\end{abstract}

\maketitle


\section{Introduction}

The properties of strong interaction matter as described by quantum
chromodynamics (QCD) at high densities and temperatures can be studied in the
laboratory by relativistic heavy-ion collision experiments. Manifestations of
this extreme state of matter created in the laboratory can be found in the
early universe, neutron star mergers and
core-collapse supernova explosions.

The processes which are able to turn the collapse of a massive stars into a
supernova explosion are not fully understood yet, see e.g.\
\cite{Janka:2012wk} for a review.  A key ingredient to core-collapse supernova
simulations is the nuclear equation of state at nonzero density, temperature
and proton fraction. During the supernova evolution high temperatures and
densities can be reached allowing for the opportunity to explore unknown
regimes of the phase diagram of strong interactions, i.e.\ the QCD phase
diagram. The conditions might be such that a new phase emerges in the core of
the collapsed star.

A possible phase transition during the supernova evolution has been studied in
\cite{Kampfer:1983zz} for a pion-condensed state and in
\cite{Takahara:1985,Takahara:1987zq} for a first-order phase transition from
hadronic matter to quark matter which can influence the supernova dynamics
such that a delayed explosion can take place. Quark matter could also appear
during the later proto-neutron star evolution as studied in
\cite{Pons:2001ar}.  The presence of a new quark matter phase during the
supernova stage has been studied in more detail in
\cite{Sagert:2008ka,Fischer:2010zz} by including effects from neutrinos. If a
new phase is present early in the evolution of the supernova, it can produce a
second shock wave with an accompanying measurable second neutrino burst
\cite{Dasgupta:2009yj}. In certain cases, different paths in the phase diagram
of QCD can be sweeped out by the delayed collapse to a black hole during the
evolution of a core-collapse supernova \cite{Ohnishi:2011jv}.  The adopted
equations of state (EoS) used above are hybrid models with a low-density
nucleonic equation of state and the simple MIT bag model extended to nonzero
temperatures as the high-density part. The merger of pure quark stars (strange
stars) was also simulated within the MIT bag model in \cite{Bauswein:2008gx}
showing distinctly different features compared to ordinary neutron star
mergers \cite{Bauswein:2009im}.

However, it is known that the MIT bag model fails in describing lattice data,
see e.g.\ \cite{Fraga:2013qra}, and is not suited to describe profound features
of dense matter QCD, as chiral symmetry restoration at high densities.  As
perturbation theory breaks down on the scale of interest here and results from
lattice QCD at high densities are not available yet, improved effective models
have to be utilized to study the regime of the QCD plasma relevant for
astrophysical applications, as core-collapse supernovae, which we focus on in
the following. The Nambu-Jona-Lasinio (NJL) model, as a chiral effective model
of QCD, has been studied for nonzero temperature and neutrino chemical
potential, relevant for proto-neutron stars, in
\cite{Steiner:2002gx,Ruster:2005ib,Laporta:2005be,Sandin:2007zr}. First exploratory
investigations of supernova explosions, which require a given electron to
baryon number ratio $Y_e$, were only undertaken recently within chiral
approaches of QCD in \cite{Fischer:2011zj}. Here the Polyakov-loop extended
version of the NJL model, the PNJL model, was used and compared to the MIT bag
model. It turned out that there are generic differences between the two model
descriptions of relevance for the supernova dynamics.

In this work the linear sigma model \cite{GellMann:1960np} is adopted as an
effective chiral model of QCD to study supernova matter, where the fundamental
particles are quarks interacting via scalar and vector meson exchange. The
quark masses are generated by nonvanishing vacuum expectation values of the
scalar fields which act as chiral condensates and model the spontaneous
breaking of chiral symmetry as observed in the vacuum of QCD. At high
temperature and/or densities the condensates melt away and the theory becomes
approximately chirally invariant. The order of the phase transition depends on
the choice of the parameters used. A detailed study is performed by varying
the different parameters of the model to delineate their role for the
properties of supernova material. A comparison between the two flavor model,
involving only the light up and down quarks, and the three flavor model, where
also the strange quark is taken into account, is performed. By using standard
methods of finite temperature field theory, the equation of state (EoS) is
calculated, which can serve as an input for core-collapse supernova and
neutron star merger simulations.


\section{Extended Linear Sigma Model}


\subsection{Standard Quark-Meson Model}

The Lagrangian of the linear sigma model with $N_f$ flavors is given by
(see e.g.\ \cite{Schaefer:2008hk}):
\begin{align}
\mathcal{L} &= \bar{\psi } ( i {\not}\partial - g \phi  ) \psi +
\mathcal{L}_{N_f} 
\label{L_int}\\
\phi &= T_a \phi _a = T_a ( \sigma _a + i \gamma _5 \pi _a ).
\end{align}
The quark spinors are of the form $\psi =(u,d) $ for $N_f =2 $ flavors and
$\psi =(u,d,s)$ for $N_f =3 $ and involve also the color degrees of
freedom $ N_c =3 $.  The Yukawa coupling between quarks and members of the
scalar/pseudoscalar meson nonets $\sigma _a / \pi _a $ is controlled by the
scalar coupling constant $g$. The generators of the underlying $U(N_f) $
symmetry group are denoted by $T_a = \lambda _a /2 $ with $a=0,..,N_f^2-1
$, which are the Pauli matrices or the Gell-Mann matrices.  The pure mesonic
contributions $ \mathcal{L} _{N_f} $ for $N_f$ flavors are of the form
\cite{GellMann:1960np,Lenaghan:2000ey}.
\begin{align}
\mathcal{L}_2 = &\frac{1}{2} \partial _{\mu } \sigma \partial ^{\mu } \sigma +
\frac{1}{2} \partial _{\mu } \vec{\pi } \partial ^{\mu }  \vec{\pi
} \label{L_2}\\ 
&- \frac{\lambda }{4} ((\vec{\pi }^2 + \sigma ^2 ) - v^2 )^2 + h \sigma \notag \\
\mathcal{L}_3 = &\Tr (\partial _{\mu } \phi ^{\dagger } \partial ^{\mu } \phi) - m^2 \Tr(\phi ^{\dagger } \phi) \label{L_3} \\
&- \lambda _1 [\Tr(\phi ^{\dagger } \phi)]^2 - \lambda _2 \Tr(\phi ^{\dagger } \phi )^2 \notag \\ 
& + c (\det(\phi ) + \det (\phi ^{\dagger } ) ) + \Tr [ H(\phi + \phi ^{\dagger } ) ] \notag
\end{align}
The first part of eq.~\eqref{L_2} is the standard kinetic term for the mesons, while
the potential part involves the Mexican hat potential, which is invariant
under chiral transformations and leads to spontaneous symmetry breaking,
as well as an explicit symmetry breaking term, controlled by the parameter
$h$. These symmetry breaking terms generate a finite vacuum expectation value
(VEV) for the $\sigma $ meson and thus a finite quark mass through
\eqref{L_int}.

The parameter $ v $ is determined by the requirement, that the minimum of
the potential lies at the pion decay constant $f_{\pi } $ and the fact that
the VEV of the pion field has to vanish due to its parity. From this follows
\begin{align}
v^{2}=f_{\pi}^{2}-\frac{h}{\lambda f_{\pi}}.
\end{align}
The masses of the $ \pi $ and $ \sigma $ mesons in the ground state are given by the second
derivatives of the potential with respect to the fields evaluated
in the vacuum at $T=0$ which fixes
\begin{align}
h & =f_{\pi}m_{\pi}^{2}\label{eq:c_fixing}\\
\lambda & =\frac{m_{\sigma}^{2}-m_{\pi}^{2}}{2f_{\pi}^{2}}.\label{eq:lambda_fixing}
\end{align}
The pion mass is directly related to the explicit symmetry breaking term and
for $ h=0 $ the pions become the massless Goldstone bosons. The light quark
masses are given according to
\begin{equation}
m_{l}=\frac{g}{2}\langle\sigma\rangle.\label{eq:g_fixing_SU2}
\end{equation}
The Lagrangian \eqref{L_3} for three flavors contains also a
standard kinetic term, as well as a meson mass term. Additionally, there are
quartic interaction terms involving the coupling constants $\lambda _1$ and
$ \lambda _2 $. The determinant term is a cubic coupling term between the
fields, which corresponds to the $U(1)_A $ anomaly. The last part is again
an explicit symmetry breaking term, with $ H= h_a T_a $.  Analogously to the
\ensuremath{SU(2)} case, through the spontaneous symmetry breaking, the
\ensuremath{\sigma} fields get nonvanishing VEVs, while again the VEVs of the
$ \pi $ fields have to vanish due to their pseudoscalar character. Following
\cite{Lenaghan:2000ey} we introduce the notation
\begin{equation}
\langle\phi\rangle\equiv\bar{\phi}\equiv T_{a}\bar{\sigma}_{a}.
\end{equation} 
Shifting $ \phi $ by their VEVs and taking into account, that only $
\bar{\sigma}_0 ,$ $\bar{\sigma} _3 $ and $\bar{ \sigma} _8 $ do not
vanish, because they respect the quantum numbers of the vacuum, the potential
takes the form \cite{Stiele:2013pma}
\begin{multline}
	U(\bar{\sigma}_{0},\bar{\sigma}_{3},\bar{\sigma}_{8})=
        m^{2}\left(\frac{\bar{\sigma}_{0}^{2}+\bar{\sigma}_{3}^{2}+\bar{\sigma}_{8}^{2}}{2}\right)
        \\ 
	-c\left(\frac{\bar{\sigma}_{0}^{3}}{3\sqrt{6}}-\frac{\bar{\sigma}_{8}^{3}}{6\sqrt{3}}
-\frac{\bar{\sigma}_{0}\bar{\sigma}_{3}^{2}}{2\sqrt{6}}-\frac{\bar{\sigma}_{0}\bar{\sigma}_{8}^{2}}{2\sqrt{6}}
+\frac{\bar{\sigma}_{3}^{2}\bar{\sigma}_{8}}{2\sqrt{3}}\right)
        \\ 
	+\lambda_{1}\left(\frac{\bar{\sigma}_{0}^{4}+\bar{\sigma}_{3}^{4}+\bar{\sigma}_{8}^{4}}{4}
+\frac{\bar{\sigma}_{0}^{2}\bar{\sigma}_{3}^{2}+\bar{\sigma}_{0}^{2}\bar{\sigma}_{8}^{2}+\bar{\sigma}_{3}^{2}\bar{\sigma}_{8}^{2}}{2}
\right)\label{VEV_potential}\\
	+\lambda_{2}\left(\frac{\bar{\sigma}_{0}^{4}}{12}+\frac{\bar{\sigma}_{3}^{4}+\bar{\sigma}_{8}^{4}}{8}
+\frac{\bar{\sigma}_{0}^{2}\bar{\sigma}_{3}^{2}+\bar{\sigma}_{0}^{2}\bar{\sigma}_{8}^{2}}{2}
        \right. \\ 
	\left. +\frac{\bar{\sigma}_{3}^{2}\bar{\sigma}_{8}^{2}}{4}+\frac{\bar{\sigma}_{0}\bar{\sigma}_{3}^{2}\bar{\sigma}_{8}}
{\sqrt{2}}-\frac{\bar{\sigma}_{0}\bar{\sigma}_{8}^{3}}{3\sqrt{2}}\right) \\
	\left.-h_{0}\bar{\sigma}_{0}-h_{3}\bar{\sigma}_{3}-h_{8}\bar{\sigma}_{8}.\right.
\end{multline}
The \ensuremath{\sigma} fields are identified with the known scalar mesons and
the \ensuremath{\pi} fields with the pseudoscalar ones
\begin{align}
(\sigma_{0},...,\sigma_{8})^{\textnormal{T}} &
\widehat{=}(\sigma_{0},a_{0}^{+},a_{0}^{-},\sigma_{3},\kappa^{+},\kappa^{-},\kappa^{0},\bar{\kappa}^{0},\sigma_{8})^{\textnormal{T}}\\ 
(\pi_{0},...,\pi_{8})^{\textnormal{T}} & \widehat{=}(\pi_{0},\pi^{+},\pi^{-},\pi_{3},K^{+},K^{-},K^{0},\bar{K}^{0},\pi_{8})^{\textnormal{T}}.
\end{align}
Their masses can be computed analogously to the $SU(2) $ case from the
potential part of the Lagrangian \eqref{L_3} through \cite{Lenaghan:2000ey} 
\begin{align}
(m_{S}^{2})_{ab} & =\frac{\partial^{2}U(\sigma,\pi)}{\partial\sigma_{a}\partial\sigma_{b}}\Big|_{\bar{\sigma}}\\
(m_{P}^{2})_{ab} & =\frac{\partial^{2}U(\sigma,\pi)}{\partial\pi_{a}\partial\pi_{b}}\Big|_{\bar{\sigma}}.
\end{align}
The condensate $ \sigma_3 $ breaks isospin symmetry explicitly. It is assumed
that the vacuum at zero temperature is isospin symmetric, so that there holds
$\sigma _3 =0 $. Thus, the explicit symmetry breaking term has to vanish,
$h_3=0$.  The quark masses are given by
\begin{align}
m_{u} &
=g\left(\frac{1}{\sqrt{6}}\bar{\sigma}_{0}+\frac{1}{2}\bar{\sigma}_{3}+\frac{1}{2\sqrt{3}}\bar{\sigma}_{8}\right)\nonumber
\\ 
m_{d} &
=g\left(\frac{1}{\sqrt{6}}\bar{\sigma}_{0}-\frac{1}{2}\bar{\sigma}_{3}
+\frac{1}{2\sqrt{3}}\bar{\sigma}_{8}\right)\label{eq:quark_masses_SU3_sigma}\\ 
m_{s} & =g\left(\frac{1}{\sqrt{6}}\bar{\sigma}_{0}-\frac{1}{\sqrt{3}}\bar{\sigma}_{8}\right).\nonumber 
\end{align}
The remaining parameters can then be determined as in refs.~\cite{Lenaghan:2000ey,Schaefer:2008hk}.

\subsection{Extension to Vector Mesons}

The standard linear sigma model is extended by a vector meson
contribution. The interaction of quarks through scalar mesons is attractive
while a repulsive force can be provided by the inclusion of vector
mesons. The Lagrangian of the vector mesons is given by
\cite{Ferroni:2010xf}
\begin{align}
\mathcal{L}_{V}=-\frac{1}{4}F^{\mu\nu}F_{\mu\nu}+\frac{\eta_{v}^{2}}{2}V^{a\mu}V_{\mu}^{a}-g_{V}^{a}\bar{\psi}\gamma^{\mu}T^{a}\psi
V_{\mu}^{a}.
\label{vector_lag}
\end{align}
where \ensuremath{V_{\mu}^{a}} are the vector meson fields. Their number
depends on the considered flavors, which are
\begin{align}
a=0,...,3\textnormal{ for }SU(2)\\
a=0,...,8\textnormal{ for }SU(3) & .
\end{align}
The \ensuremath{T^{a}} are again the generators of the appropriate groups. The
Lorentz index \ensuremath{\mu} indicates the vector character of the
mesons. The first term in equation \eqref{vector_lag} is the standard kinetic
term for vector particles, while the second one is the mass term. The vector
mesons are coupled to the quark fields by a Yukawa type interaction term.

\subsection{Grand Canonical Potential}

Assuming local equilibrium allows to work in a mean field approximation. The
meson fields are treated as classical fields. As pointed out before, the VEVs
of the pion fields have to vanish due to parity while the $\sigma$ fields adopt
a finite value. In the case of three flavors a finite VEV is only investigated
for the \ensuremath{\sigma_{0}}, \ensuremath{\sigma_{3}} and
\ensuremath{\sigma_{8}} fields and it is assumed for the others to
vanish. Furthermore, the spatial components of the vector meson fields have to
vanish because of rotational symmetry. Additionally, for two flavors only the
zeroth and third flavor group component do not vanish because of isospin
invariance. For three flavors one ends up with an additional vector meson
field from the eighth component of the vector field nonet. The mathematical
vector fields can be directly identified with the physical ones. For $ N_f = 2
$ one finds
\begin{align}
\omega & \widehat{=}\bar{V}_{0}^{0}\nonumber \\
\rho & \widehat{=}\bar{V}_{0}^{3}.
\end{align}
For $ N_f = 3 $ one decouples the strange quark sector by assuming ideal
mixing and one finds
\begin{align}
\omega & \widehat{=}\sqrt{\frac{2}{3}}\bar{V}_{0}^{0}+\frac{1}{\sqrt{3}}\bar{V}_{0}^{8}\nonumber \\
\rho & \widehat{=}\bar{V}_{0}^{3}\\
\phi & \widehat{=}\frac{1}{\sqrt{3}}\bar{V}_{0}^{0}-\sqrt{\frac{2}{3}}\bar{V}_{0}^{8}.\nonumber 
\end{align}
The rotation of fields into each other results in the change of the effective
coupling constants according to
\begin{equation}
\frac{g_V}{2}=g_{\omega } = g_{\rho } = \frac{g_{\phi}}{\sqrt{2}}
\end{equation}
as dictated by $SU(3)$ symmetry and ideal mixing. The fields occur in
combination with the Gell-Mann matrices, that is why also the physical fields
have to be given with respect to a basis which is defined as
\begin{equation}
\chi^{\omega}=\left(\begin{array}{ccc}
1 & 0 & 0\\
0 & 1 & 0\\
0 & 0 & 0
\end{array}\right)\;\chi^{\rho}=\left(\begin{array}{ccc}
1 & 0 & 0\\
0 & -1 & 0\\
0 & 0 & 0
\end{array}\right)\;\chi^{\phi}=\left(\begin{array}{ccc}
0 & 0 & 0\\
0 & 0 & 0\\
0 & 0 & 1
\end{array}\right).
\end{equation}
The $SU(2) $ mean field Lagrangian is then given by
\begin{align}
\mathcal{L}= & \bar{\psi}\left( i {\not}\partial+ \mu \gamma ^0
  -\frac{g}{2}\sigma-g_{\omega}\gamma^{0}\omega-g_{\rho}\gamma^{0}\tau^{3}\rho\right)\psi
\label{SU2_mean_field_int}\\  
 -&
 \frac{\lambda}{4}(\sigma^{2}-v^{2})^{2}
+ h\sigma+\frac{m_{\omega}^{2}}{2}\omega^{2}+\frac{m_{\rho}^{2}}{2}\rho^{2}
\label{SU2_mean_field_pot} 
\end{align}
Here also the chemical potential was added. It is a diagonal matrix with the
entries $ \mu _u $ and $ \mu _d $. The vector meson contributions are of
the same form and one defines an effective chemical potential
\begin{equation}
\tilde{\mu}=\left(\begin{array}{cc}
\tilde{\mu}_{u} & 0 \\
0 & \tilde{\mu}_{d} 
\end{array}\right)
=\left(\begin{array}{cc}
\mu_{u}-g_{\omega}\omega-g_{\rho}\rho & 0 \\
0 & \mu_{d}-g_{\omega}\omega+g_{\rho}\rho \end{array}\right).
\end{equation}
Now we turn to the case of SU(3) flavor symmetry.
Analog to the vector meson fields, also the condensates are rotated to
decouple the strange and nonstrange sectors
\cite{Lenaghan:2000ey,Schaefer:2008hk}
\begin{equation}
\left(\begin{array}{c}
\sigma_{x}\\
\sigma_{y}
\end{array}\right)=\frac{1}{\sqrt{3}}\left(\begin{array}{cc}
\sqrt{2} & 1\\
1 & -\sqrt{2}
\end{array}\right)\left(\begin{array}{c}
\sigma_{0}\\
\sigma_{8}
\end{array}\right)\label{eq:transf_matrix}
\end{equation}
where $\sigma_{x}$ is called the nonstrange condensate and $\sigma_{y}$ the
strange condensate. Using this new notation the meson potential is rewritten
and the $SU(3) $ mean field Lagrangian takes the form
\begin{align}
\mathcal{L}= & \bar{\psi}\left[ i {\not}\partial+ \mu \gamma ^0 -
  g\left(\sigma_{0}\frac{\lambda^{0}}{2}+\sigma_{3}\frac{\lambda^{3}}{2}+\sigma_{8}\frac{\lambda^{8}}{2}\right)\right.\\ 
 &
 \left.-g_{\omega}\gamma^{0}\omega\chi^{\omega}-g_{\rho}\gamma^{0}\rho\chi^{\rho}-g_{\phi}\gamma^{0}\phi\chi^{\phi}\right]\psi
 \nonumber \\ 
-&
\frac{m^{2}}{2}(\sigma_{x}^{2}+\sigma_{y}^{2}+\sigma_{3}^{2})+\frac{c}{2\sqrt{2}}(\sigma_{x}^{2}\sigma_{y}-\sigma_{3}^{2}\sigma_{y})
\nonumber
\\ 
 &
 -\frac{1}{8}\left(2\lambda_{1}+\lambda_{2}\right)\left(\sigma_{x}^{4}+\sigma_{3}^{4}\right)
-\frac{1}{4}\left(\lambda_{1}+\lambda_{2}\right)\sigma_{y}^{4}\nonumber 
 \\ 
- &
\frac{\lambda_{1}}{2}\left(\sigma_{x}^{2}\sigma_{y}^{2}+\sigma_{3}^{2}\sigma_{y}^{2}+\sigma_{x}^{2}\sigma_{3}^{2}\right)
-\frac{3}{4}\lambda_{2}\sigma_{x}^{2}\sigma_{3}^{2}\nonumber 
\\ 
+ &
h_{x}\sigma_{x}+h_{y}\sigma_{y}+\frac{m_{\omega}^{2}}{2}\omega^{2}+\frac{m_{\rho}^{2}}{2}\rho^{2}+\frac{m_{\phi}^{2}}{2}\phi^{2}.\nonumber 
\end{align}
The chemical potential gets an additional entry for the strange quark of the
form $ \tilde{\mu }_s = \mu _s - g_{\phi } \phi $.  The grand canonical
potential is connected to the Lagrangian through the path integral over the
quark fields, which are the only remaining quantum fields
\begin{align}
\Omega &= - \frac{T}{V} \ln \mathcal{Z} \\
\mathcal{Z}&=\int\mathcal{D}\bar{\psi}\mathcal{D}\psi\exp\left[\int_{0}^{\beta}\intd\tau\int\intd^{3}
  x\, \mathcal{L}(\bar{\psi},\psi)\right]. 
\end{align}
with the temperature $T $ and $\beta= {1}/{T}$. The electron contribution is
omitted here, since it is fully decoupled and can be computed straight
forward. The interesting thermodynamic quantities can be derived from the
grand canonical potential through the standard equations from statistical
physics, 
\begin{eqnarray}
&&p = -\Omega\;,\quad s = -\left.\frac{\partial\Omega}{\partial
    T}\right|_{\mu_i=\text{const}},\quad n_i =
-\left.\frac{\partial\Omega}{\partial\mu_i}\right|_{T=\text{const}}\\ 
&&\epsilon  = Ts+\Omega+\sum_i\mu_i n_i\;.
\end{eqnarray}
The values of the condensates and vector meson fields are computed by solving
the gap equations
\begin{align}
  SU(2):\; &
  \frac{\partial\Omega}{\partial\sigma}=\frac{\partial\Omega}{\partial\omega}=\frac{\partial\Omega}{\partial\rho}=0
\label{gap_eq_su2}\\
  SU(3):\; &
  \frac{\partial\Omega}{\partial\sigma_{x}}=\frac{\partial\Omega}{\partial\sigma_{y}}=\frac{\partial\Omega}{\partial\sigma_{3}}
=\frac{\partial\Omega}{\partial\omega}=\frac{\partial\Omega}{\partial\rho}=\frac{\partial\Omega}{\partial\text{\ensuremath{\phi}}}
=0 
\label{gap_eq_su3}  
\end{align}
in the mean field approximation.

\subsection{Parameter Fixing}

As described before, the parameters of the potentials are fixed by using
measured meson masses and decay constants given by the Particle Data Group
\cite{Beringer:1900zz}. For the $SU(2) $ case it is used
\begin{alignat}{1}
\begin{array}{cc}
\begin{aligned}m_{\pi^{\pm}} & =140\,\textnormal{MeV}\\
m_{\omega} & =783\,\textnormal{MeV}
\end{aligned}
 & \begin{aligned}m_{\pi^{0}} & =135\,\textnormal{MeV}\\
m_{\rho} & =775\,\textnormal{MeV}
\end{aligned}
\end{array}\label{eq:meson_mass_SU2}
\end{alignat}
with $f_{\pi} =92$~MeV.
Since we do not distinguish between the different pions, an averaged value is used.
In the $SU(3) $ case more values need to be fixed, additionally 
\begin{alignat}{1}
\begin{array}{cc}
\begin{alignedat}{1}m_{K^{\pm}} & =494\,\textnormal{MeV}\\
m_{\eta} & =548\,\textnormal{MeV}\\
f_{K} & =110\,\textnormal{MeV}
\end{alignedat}
 & \begin{alignedat}{1}m_{K^{0}} & =498\,\textnormal{MeV}\\
m_{\eta'} & =958\,\textnormal{MeV}\\
m_{\phi} & =1019\,\textnormal{MeV}
\end{alignedat}
\end{array}
\end{alignat}
are used, where again the kaon masses are averaged. The scalar coupling
constant $ g $ is fixed by the constituent light quark mass  $ m_l = 300 \,
\textnormal{MeV} $. 
 
To solve the equations \eqref{gap_eq_su2} or \eqref{gap_eq_su3} also the
temperature and chemical potentials have to be known. They are computed by
implementing constraints, corresponding to supernova matter. The baryon
density is fixed and given by
\begin{align}
n_{B}= & \frac{n_{u}+n_{d}+n_{s}}{3}.
\end{align}
As further input in supernovae simulations the electron-baryon ratio $Y_{e}$
is fixed according to
\begin{align}
Y_{e}\equiv\frac{n_{e}}{n_{B}} & =\frac{3n_{e}}{n_{u}+n_{d}+n_{s}}.\label{eq:e_B_rate}
\end{align}
As a standard value $Y_{e}=0.2$ is used for the typical value at core bounce
of a supernova, following \cite{Fischer:2011zj}.
Demanding electric charge neutrality yields the condition
\begin{align}
0= & q_{u}n_{u}+q_{d}n_{d}+q_{s}n_{s}+q_{e}n_{e}.
\end{align}
with the corresponding charges $q_i$ of the quarks and the electron.
Additionally, local equilibrium with respect to the weak interaction process $
s+u \leftrightarrow d+u $ is assumed for three flavors, from which follows 
that $\mu _d = \mu_s $.

During the supernova explosion temperatures of several tens of MeV are reached
\cite{Fischer:2010wp}.  The entropy gives a constraint to determine the
temperature relevant for supernova explosions. For the variation limits of the
entropy we use \cite{Fischer:2010wp} as guidance. So we choose a standard
value of one $k_B$ per baryon and explore the interval of $0.5-4\, k_B /
\textnormal{baryon}$.

Not all parameters can be fixed, the remaining ones are varied in the further
analysis. For the vector coupling a range of $ g_{\omega } = (1-10) $ is
used, since it should be in the same range as the scalar vector coupling $
g$. As a standard value we choose $ g_{\omega } = 3.0 $. 

The mass of the $ \sigma $ meson is not determined well experimentally yet.
Recently, it has been identified with the resonance $f_{0}(500)$
\cite{Beringer:1900zz} that has replaced the broad $f_{0}(600)$ resonance.
Other approaches assign the $f_{0}(1370)$ resonance with the $\sigma$ meson
\cite{Parganlija:2012fy}. Hence, it will be varied in the range of
$(400-1000)\:\textnormal{MeV}$, but the standard value
$m_{\sigma}=600\:\textnormal{MeV}$ is used \cite{Schaefer:2008hk}. The upper
bound is determined by the fact that the model saturates around
$m_{\sigma}\approx1100\,\textnormal{MeV}$ \cite{Schaefer:2008hk}.

To study the influence of the electron-baryon-rate, $ Y_e $ will also be
varied in the range $(0.0-0.5) $.


\section{Phase Diagram}

\begin{figure}
\centerline{\includegraphics[width=0.5\textwidth]{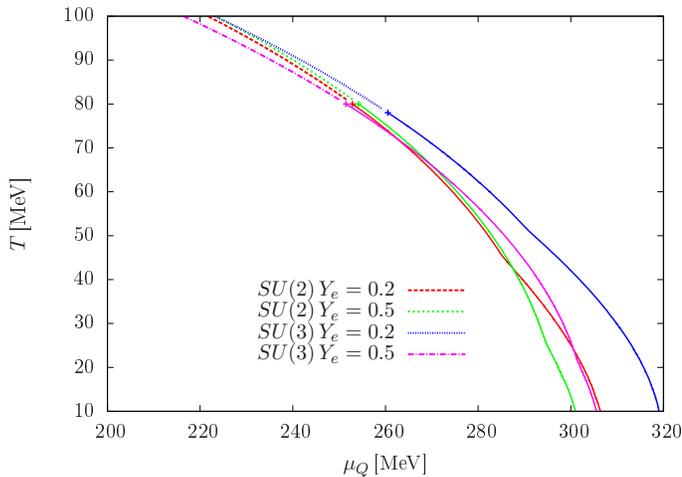}} 	
\caption{\label{fig:phase_diag_mu}Phase diagram in $\mu_{Q}-T$ space for the
  different flavor cases, with the standard $Y_{e}$ values, $g_{\omega}=3.0,$
  $m_{\sigma}=600\,\textnormal{MeV}.$}
\end{figure}

In figure \ref{fig:phase_diag_mu} the phase diagram for the $SU(2)$ and
$SU(3)$ flavor case for $Y_{e}=0.2$ and $Y_{e}=0.5$ is shown.  A first order
phase transition is observed for all cases at low temperatures represented by
the solid lines. This phase transition is associated with the approximate
restoration of chiral symmetry in the light quark sector where the $\sigma_x$
field serves as an corresponding order parameter.

For both flavor cases a critical endpoint is observed which lies approximately
at the same temperature $(T\approx80\,\textnormal{MeV})$. For higher
temperatures the phase transition is a crossover, i.e.\ the system goes
smoothly from one phase into the other so that no jump in the order parameter
appears and no mixed phase of the low- and high-density phase is present. The
phase transition lines for the different cases shown are quite close to each other.
However, one recognizes that for a lower electron fraction the phase transition happens
at lower quark chemical potential for the three flavor case.  The phase
transition line for $N_f=3$ is located at higher quark chemical potential
compared to $N_f=2$ for $Y_e=0.2$. 

The location of the transition line is similar to the findings in
\cite{Schaefer:2008hk} for the QM model. Small differences are present even
for the isospin symmetric case and should account for the fact, that effects
from vector meson exchange are considered here, which are expected to be of
more relevance for the high-density moderate temperature region of the QCD
phase diagram of interest in the present study.

In the following the change of the transition line due to
changes in the parameters will be illustrated in more detail.

To understand the phase transition structure, the evolution of the relevant
scalar fields shall be discussed. We find that only the nonstrange condensate
$\sigma_{x}$ shows a strong jump across the phase transition line and reaches
values close to the one of a chirally fully restored phase. Although a small
jump in the strange scalar field $\sigma_{y}$ is observed, its value stays
high (over $0.5$) and so the strange quarks remain massive. This behavior is
expected, as electric charge neutrality and the electron fraction are
implemented. Since the up quark is the only remaining positively charged
particle and down and strange chemical potentials are the same, the density of
the strange quark has to be suppressed by a large mass term, represented by a
large value of $\sigma_{y}$. Following this, for lower values of $Y_{e}$ the
strange quark condensate melts away at lower densities, as a higher number
density of strange quarks is favored.

The value of the third scalar field, $\sigma_{3}$, relevant for isospin breaking
effects, stays always low, so that the mass difference between up and down
quark remains small.

\begin{figure}
\centerline{\includegraphics[width=0.5\textwidth]{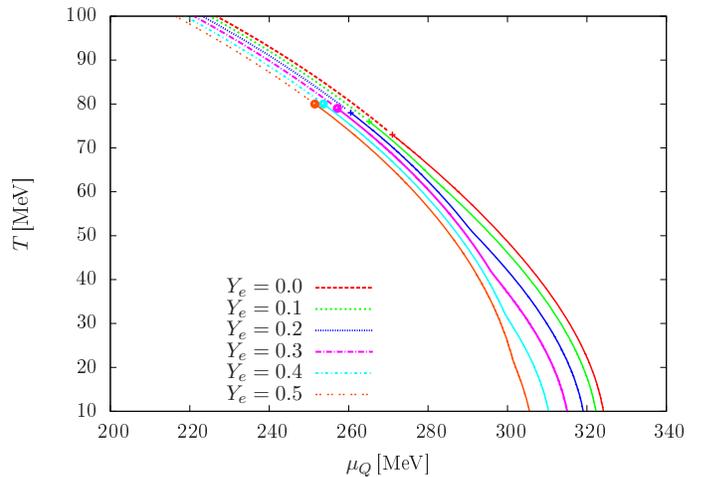}} 	
\caption{\label{fig:ye_mu_SU3}$SU(3)$ phase diagram in $\mu_{Q}-T$ space for
  different $Y_{e}$ with $g_{\omega}=3.0,$ $m_{\sigma}=600\,\textnormal{MeV}.$
  The phase transition line is shifted to lower values of $\mu_{Q}$ for
  increasing $Y_{e}.$}
\end{figure}

Figure \ref{fig:ye_mu_SU3} shows how the electron fraction influences the
phase diagram for three quark flavors. The phase transition takes place at
lower quark chemical potential for an increased electron fraction. For a
higher value of $Y_{e}$ the strange quark gets suppressed, because of charge
neutrality, and so the light quark density is higher leading to a larger
contribution in the gap equation and an onset of the chiral phase transition
at lower densities.  For the $SU(2) $ case the system is not as sensitive to
changes in $Y_{e}$ as for the three flavor case, since the strange quarks are
not available to be replaced by the light quarks.

\begin{figure}
\centerline{\includegraphics[width=0.5\textwidth]{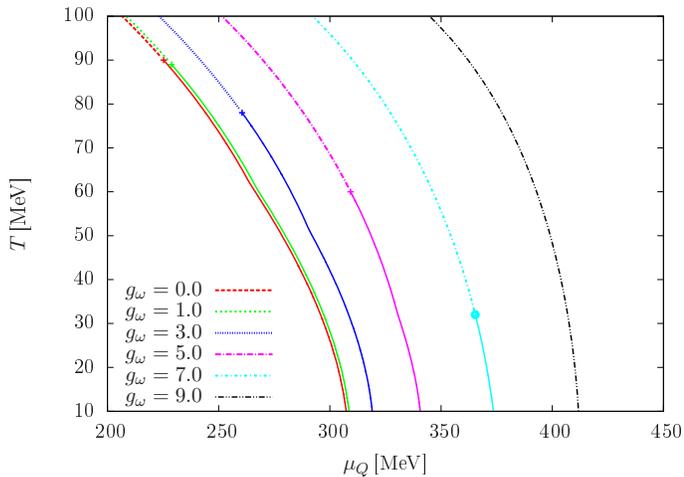}}
\caption{\label{fig:g_y2_su3}$SU(3)$ phase diagram in $\mu_{Q}-T$ space for
  $Y_{e}=0.2,$ $m_{\sigma}=600\,\textnormal{MeV}$ and different vector
  coupling constants $g_{\omega}.$ The critical endpoint moves to smaller
  temperatures for an increase in $g_{\omega}$ until the phase transition is a
  crossover over the whole plane. The transition line moves to larger
  $\mu_{Q}$ for growing $g_{\omega}.$}
\end{figure}
  
The dependence of the phase transition line on the vector coupling
$g_{\omega}$ is shown in figure \ref{fig:g_y2_su3}. The phase transition gets
smoother with increasing vector coupling constant because of the repulsive
character of the vector mesons. Thus, the jumps in the condensates become
smaller, until the transition is a crossover at all considered
temperatures. The transition line also moves to higher quark chemical
potential $\mu_{Q}$, since the effective chemical potentials are lowered as
the vector meson contribution increases with increasing
$g_{\omega}$. Therefore a larger $\mu_{Q}$ is needed for the phase transition
to happen. Our findings are in line with the results in the Nambu-Jona-Lasinio
model with vector interactions, see e.g.\ Ref.~\cite{Sasaki:2006ws,Fukushima:2008wg}.

The critical endpoint is located at lower temperature for increasing values of
$g_{\omega}$, as it is expected from the discussion above, but it varies only
slightly between the different flavor and $Y_{e}$ cases. For $g_{\omega}=9.0$
the critical endpoint vanishes from the phase diagram, so that the phase
transition is always a crossover.  Additionally, it is observed that the
sensitivity to changes in $g_{\omega}$ is much higher compared to the changes
in $Y_{e}$.

\begin{figure}
\centerline{\includegraphics[width=0.5\textwidth]{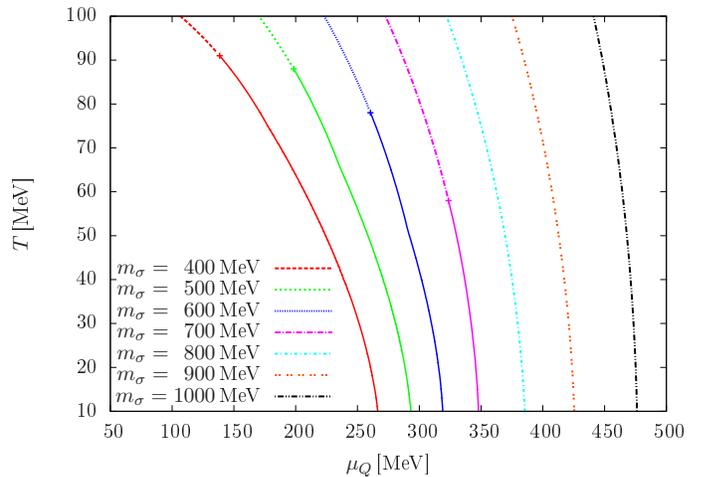}}
\caption{\label{fig:ms_y2_su3}$SU(3)$ phase diagram in $\mu_{Q}-T$ space for
  $Y_{e}=0.2,$ $g_{\omega}=3.0$ and different scalar meson masses
  $m_{\sigma}.$ The critical endpoint moves to lower temperatures for
  increasing $m_{\sigma},$ until for all temperatures a crossover is
  observed. The transition line moves to larger $\mu_{Q}$ for higher
  values of $m_{\sigma}$.}
\end{figure}

Figure \ref{fig:ms_y2_su3} illustrates the changes of the phase diagram with
the mass of the $\sigma$ meson $m_{\sigma}$. For very high $\sigma$ masses the
phase transition does not take place until very high chemical potentials are
reached, as also seen in Ref.~\cite{Schaefer:2008hk}.  The role of $m_{\sigma}$ can
be understood more easily within the $SU(2)$ model. The $\sigma$ mass fixes
the $\lambda$ parameter in the meson potential. Thus for an increasing
$m_{\sigma}$ also $\lambda$ increases and the potential becomes
deeper. Therefore, more energy is needed to develop the second minimum and the
phase transition is shifted to higher values of $\mu_{Q}$ and becomes
smoother. For the $SU(3)$ case the situation is less transparent, since the
scalar meson mass is fixed by solving a system of nonlinear equations
\cite{Lenaghan:2000ey,Schaefer:2008hk}.  However, the dependence of the phase
transition line on $m_{\sigma } $ turns out to be the same.  For large values
of $m_{\sigma}$, the phase transition is always a crossover, as it was also
observed for large values of the vector coupling constant $g_{\omega}$, i.e.\
for a large vector repulsion between the quarks.

For a lower $\sigma$ meson mass the phase transition appears at considerably
smaller values of $\mu_{Q}$. Note, that the effect of a lower value of
$m_{\sigma}$ can be compensated by an increase in the vector coupling
$g_{\omega}$ to shift the phase transition line back to larger values of
$\mu_{Q}$. In the $m_{\sigma}$ parameter range that is considered here, the
changes of the critical quark chemical potential are similar to the case when
the vector coupling constant $g_{\omega}$ is varied within the adopted range
considered to be natural. But one should recall, that the $g_{\omega}$
parameter range was only fixed by analogy arguments, while the mass
$m_{\sigma}$ is varied according to the mass range motivated by experimental
data.


\section{Equation of State}

\begin{figure}
\centerline{\includegraphics[width=0.5\textwidth]{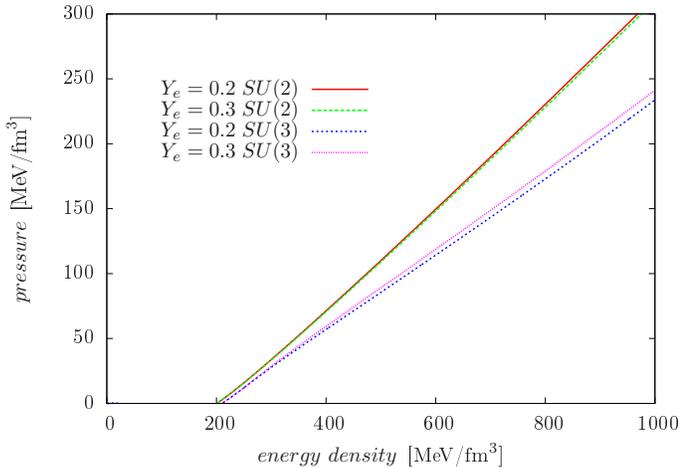}}
\caption{\label{fig:eos}Equation of state for different $Y_{e}$ and number of
  flavors with $s=1.0\, k_{B}/\textnormal{baryon},$ $g_{\omega}=3.0,$
  $m_{\sigma}=600\,\textnormal{MeV}.$ A nearly linear relation between
  pressure and energy is found. At a given energy density, the two flavor case
  provides a larger pressure compared to the three flavor one with a larger
  slope.  The nonzero energy density at vanishing pressure is due to the used
  Maxwell construction from the vacuum to the matter phase. At
  $T=0\,\textnormal{MeV}$ no quarks are present before the phase transition
  takes place.}
\end{figure}

In the following, we consider supernova matter which evolves adiabatically at
a fixed entropy per baryon $s$. In Figure \ref{fig:eos} the equation of state
is shown for a fixed value of $s=1.0\, k_{B}/\textnormal{baryon}$ for the
$SU(2)$ and $SU(3)$ flavor cases. The value of $Y_{e}=0.3$ was chosen to
compare our results with Ref.~\cite{Fischer:2011zj}.

One notices that the equation of state has a larger slope in the $SU(2)$
flavor case compared to the $SU(3)$ flavor case. The electron fraction $Y_e$
has a very little influence on the equation of state. It seems that the
computed EoS in the SU(3) case follows approximately the EoS of a free gas of
relativistic massless particles, $p={\epsilon}/{3}$, as chiral symmetry is
approximately restored. However, due to the explicit symmetry breaking terms
the particles are never really massless. Additionally, there is a contribution
from the vector mesons. Both effects combine to the observed slope of the EoS.
For two flavors the system with a smaller electron fraction has a larger
pressure, while for three flavors the pressure in the system with more
electrons is higher.  The difference between the $SU(2)$ and $SU(3)$ flavor
cases is due to the presence of the strange quark, which does not become
massless \cite{Schmitt:2010pn}.  The EoS for the $SU(3)$ case with a higher
electron fraction lies above the one with the lower electron fraction, because
the strange quark fraction is lower in the first case.

\begin{figure}
\centerline{\includegraphics[width=0.5\textwidth]{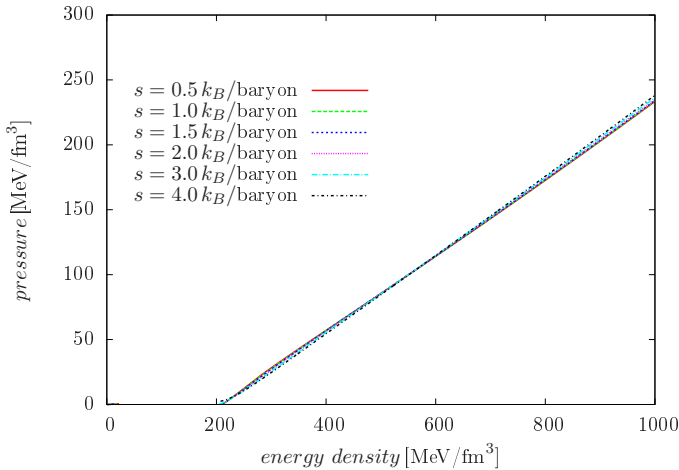}}
\caption{\label{fig:svar_y3_su3_eos}$SU(3)$ equation of state for $Y_{e}=0.2$
  and different entropy per baryon values $s$ with $g_{\omega}=3.0,$
  $m_{\sigma}=600\,\textnormal{MeV}.$ The different cases lie nearly on top of
  each other.}
\end{figure}

The effect of varying the entropy per baryon is shown in figure
\ref{fig:svar_y3_su3_eos}.  The differences are so small that the lines lie
nearly on top of each other for all values of $s$ in the given interval. This
is caused by the low temperatures involved, so that the thermal contributions
are small while the shape of the EoS is mostly determined by the chemical
potentials and the interactions.

\begin{figure}
\centerline{\includegraphics[width=0.5\textwidth]{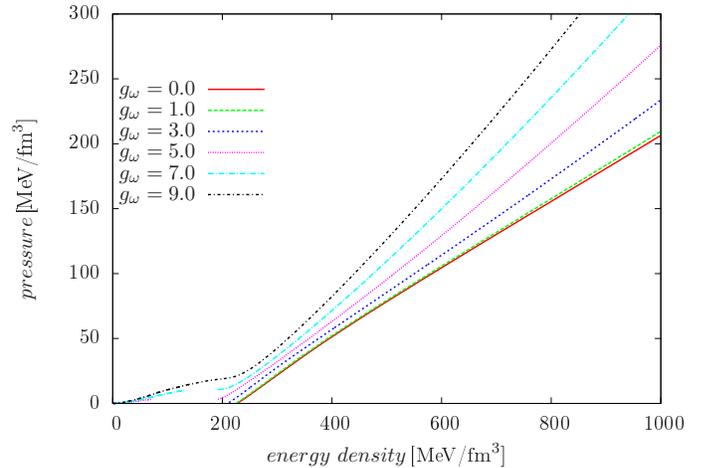}}
\caption{\label{fig:gvar_y3_su3_eos}$SU(3)$ equation of state for $Y_{e}=0.2,$
  $s=1.0\, k_{B}/\textnormal{baryon},$ $m_{\sigma}=600\,\textnormal{MeV}$ and
  different vector meson couplings $g_{\omega}.$ At small energy densities one
  recognizes that the mixed phase region gets smaller with increasing
  $g_{\omega}$. The slope of the EoS increases for larger values of $g_\omega$.}
\end{figure}

For an increasing value of $g_{\omega}$ the slopes of the equations of state
rise, as shown in Figure \ref{fig:gvar_y3_su3_eos}. This behaviour is quite
generic and well known from the relativistic $\sigma-\omega$ model for nuclear
matter \cite{Boguta:1981px,Waldhauser:1987ed} and is also seen in NJL model
calculations with vector coupling terms, see e.g.\ the discussion in
\cite{Buballa:1996tm} and Ref.~\cite{Lenzi:2012xz} for an explicit plot. In
principal the EoS would eventually reach the limit $p=\epsilon$, which would
mean that the speed of sound equals the speed of light setting the causal
limit.  This high-density limit originates from the inclusion of the vector
meson exchange, otherwise the limit would be $p=\epsilon/3$. Hence, a nonzero
vector coupling results in the wrong high-density limit as dictated by
asymptotic freedom of QCD, see \cite{Fraga:2013qra,Kurkela:2014vha} and the
discussion in \cite{Buballa:2014jta}. But at the energy densities that are
considered here the impact of the contribution from the vector mesons are not
strong enough to enter this regime. For the cases of a first order phase
transition an almost vanishing pressure for a nonvanishing energy density is
found followed by a linear increase in the pressure. This behavior is not
found if the phase transition is a crossover, as there is no jump in the
energy density.

\begin{figure}[bt]
\centerline{\includegraphics[width=0.5\textwidth]{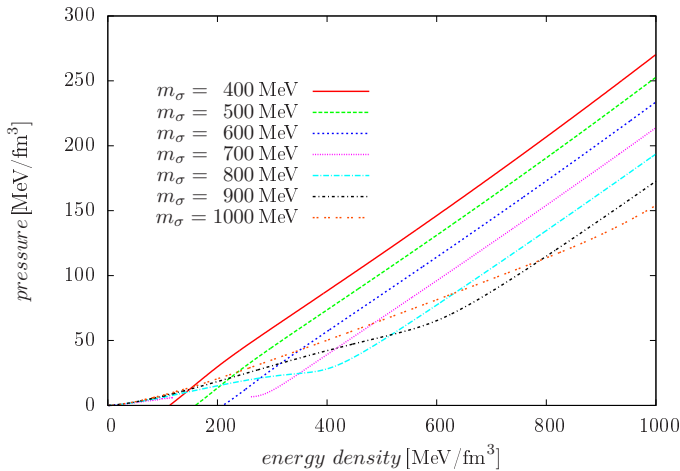}}
\caption{\label{fig:msvar_y3_su3_eos}$SU(3)$ equation of state for
  $Y_{e}=0.2,$ $s=1.0\, k_{B}/\textnormal{baryon},$ $g_{\omega}=3.0$ and
  different $\sigma$ meson masses $m_{\sigma}.$ At large energy densities the
  EoS shows the same slope and the curves are just shifted to higher energy
  densities with increasing values of $m_\sigma$. If the phase transition is a
  crossover, the lines have a smaller slope at small energy densities.}
\end{figure}

The influence of the $\sigma $ mass on the equation of state is shown in
Figure~\ref{fig:msvar_y3_su3_eos}. One notices that the impact of varying the
sigma meson mass is different for the low- and high-energy density
regions. The slope of the curves are unaffected and the EoS is just shifted
parallel in the high energy density limit when varying $m_\sigma$. For a more
massive scalar meson, the EoS has a lower pressure at the same energy density.
However, differences occur at low energy densities. If the phase transition is
a crossover, i.e.\ for high $\sigma$ meson masses, the slope of the EoS is
much smaller compared to the high-energy density limit. Here, the EoS for
higher values of $m_\sigma$ has a higher pressure at the same energy density,
contrary to the high-energy density limit.


\section{Summary} 

In this work the structure of the phase diagram and the EoS calculated from
the linear sigma model in a mean field approximation was analyzed. The model
was expanded by adding vector mesons, which give an additional contribution to
the quark chemical potentials. Calculations for the two and three flavor cases
were presented.

The parameters were fixed by measured meson masses and decay constants.
The remaining free parameter, like the mass of the $\sigma$ meson $m_{\sigma}$ and
the vector coupling constant $g_{\omega}$, were varied and their influences on
the phase diagram and the equation of state were investigated.
The conditions characteristic for core-collapse supernova explosions served
as further input. These are charge neutrality, $\beta-$equilibrium
with respect to weak interactions and a given electron-baryon fraction.

The phase diagram was analyzed and a first order phase transition at low
temperatures was observed for certain parameters.  For higher temperatures the
phase transition line ends in a critical endpoint, after which the phase
transition is a crossover. By increasing the vector meson
coupling constant the phase transition gets shifted to higher $\mu _Q$, but
for too large values the first order phase transition vanishes and a crossover
is observed for all temperatures. The same behavior was seen for increasing
$\sigma$ masses. The effect of a lowered $\sigma$ mass could be compensated by
an increased value of $g_{\omega}$.

The equation of state was calculated for a given entropy per baryon, with
typical values found in supernova simulations. The particular value for the
entropy per baryon has little influences on the EoS.  An increasing vector
repulsion, i.e.\ increasing $g_{\omega}$, leads to a higher slope of the EoS.
Higher values of $m_{\sigma}$ shift the EoS to lower pressures at constant
energy densities. However, the slope of the EoS at high pressures stays
constant, whereas at low energy densities differences occur due to the change
of the order of the phase transition.

Whether the computed equation of state is useful in a supernova explosion has
to be tested in simulations. We stress, that the model presented here lacks a
low-density hadronic phase.  Thus it should be matched at low energy densities
to a hadronic EoS for most cases, unless strange quark matter is absolutely
stable at vanishing temperature.  Recent improvements of the Polyakov-loop
extension of the quark-meson model
\cite{Haas:2013qwp,Stiele:2013pma,Herbst:2013ufa} can be considered in this
framework and the conditions of supernova matter studied here can be worked
into the investigation of the nucleation timescales of a quark phase
\cite{Mintz:2009ay,Mintz:2012mz}.

More importantly, the quark meson model has to be confronted with the observed
new mass limit for compact stars from the mass measurement of the pulsars PSR
J1614-2230 with a mass of $M=1.97\pm0.04\,\textnormal{M}_{\odot}$
\cite{Demorest:2010bx} and PSR J0348+0432 \cite{Antoniadis:2013pzd} with a
mass of $M=2.01\pm0.04\,\textnormal{M}_{\odot}$. This is work in progress and
will allow to constrain the parameter space more strictly than it was possible
in this analysis \footnote{A. Zacchi, R. Stiele, J. Schaffner-Bielich, in
  preparation (2014)}.

\begin{acknowledgments}
  This work is supported by the German Federal Ministry of Education and
  Research (BMBF) under grants FKZ 05P12VHCTG and 06HD7142, by the German
  Research Foundation (DFG) within the framework of the excellence initiative
  through the Heidelberg Graduate School of Fundamental Physics (HGSFP) and
  through the Helmholtz Graduate School for Heavy-Ion Research (HGS-HIRe) and
  the Graduate Program for Hadron and Ion Research (GP-HIR) by the
  Gesellschaft f\"ur Schwer\-ionen\-forschung (GSI), Darmstadt and the
  Alliance Program of the Helmholtz Association HA216/EMMI.
\end{acknowledgments}

\bibliography{all_new,myliterature}
\bibliographystyle{apsrev4-1.bst}

\end{document}